\documentclass[12pt,a4paper]{article}
\usepackage{graphicx}
\begin{document}
\begin{center}
\vspace{7.cm}
{\Large \bf {On the consistency of LEAR experimets and FENICE in the sector
of $p\bar p$ interaction near the threshold}} \\
\vspace{0.5cm} {\bf
{$^{(1)}$M.Majewski,$^{(2)}$G.V.Meshcheryakov,$^{(2)}$V.A.Meshcheryakov}} \\
{$^{(1)}$University of Lodz, Department of Theoretical Physics, ul.~Pomorska
149/153, 90-236 Lodz, Poland.\\ $^{(2)}$Joint Institute for Nuclear Research,
    Dubna 141980, Moscow Region, Russia.} \\ {\it
   {(1)--mimajew@mvii.uni.lodz.pl~~(2)-- mva@thsun1.jinr.dubna.su}} \\
\newpage
\vspace{1.7cm} ABSTRACT \end{center} \bigskip Some experiments on LEAR
   obtained unusual behavior of the $p\bar p$ interaction near the threshold.
   The  experiments on $p\bar p$ forvard scattering detected zeros and big
   variation of $\rho$ and at the same time a smooth rising of
   ${\sigma}_{tot}$ with lowing energy. Many models has difficulties in
 explanating this fact.  In the PS170 experiment with a good statistical
accuracy the unexpected behavior of the proton electromagnetic form factor was
found. All these experiments can be considered as an indication for the
existence of a low lying $p\bar p$ bound state 'baryonium'. This statement
coincides with that made for interpretation of the energy dependence of the
total cross-section $e^{+}e^{-}\to hadrons$ in FENICE.  There is a model
(based on analyticity) which explained aforementioned experiments and the fact
that this 'baryonium' is not seen in the OBELIX $p\bar p$ annihilation
cross-section. Thus LEAR experiments and FENICE one are consistent near
$p\bar p$ threshold and compatible with the existence of 'baryonium'.
\\[\baselineskip]
~~~~14.20---Baryons and baryons resonans(including antiparticles).
\newpage
\section{The database and previous knowledge} The experiment on LEAR which is
a part of the CERN antiproton complex gives a rich information about low
energy antiproton physics.  The experiments (PS172, PS173) \cite{1,2} on
$p\bar p$ scattering give the data on $d\sigma/d\Omega$, $\sigma_{tot}$ and
$\rho$. To search for bound state cross section measurements are the most
straightforward experiments to perform. The analysis of $d\sigma/d\Omega$
gives an indication of the bound states near $p\bar p$ threshold \cite{3}.
Some of them are consistent with the strong interaction shifts and width of
protonium \cite{4}. A resonance (a bound state having a mass bigger the $p\bar
p$ threshold) may be seen as a bump in $\sigma_{tot}$.  But the measurements
of the $p\bar p$ total cross section above 180~MeV/c indicate its smoothly varying
behaviour \cite{2}. The most remarkable result in $p\bar p$ elastic scattering
has appeared in the data on real-to-imaginary ratio of the forward scattering
amplitude $\rho$ which at LEAR was measured down to 180~MeV/c \cite{2}. For
the range $350<p_l<700~MeV/C$ the behaviour of $\rho$ can be explained by
insertion of a pole below threshold in the dispersion relation analysis
\cite{5}. But LEAR measurements \cite{2,6} below 350~MeV/c indicate that the
$\rho$ is turning upward again. The reason for this unusual behaviour is not
yet clear. It might be caused by a $p\bar p$ bound state \cite{7} but
not by an $n\bar n$ threshold \cite{8}. Experimental the $\rho$
was always determined from elastic differential cross section in the
Coulomb-nuclear interference region. The method used to extract $\rho$ from such data sometimes has been criticized \cite{9}.
But at high energies the method is consistent with the predictions of dispersion relations. So
$\rho$ from \cite{2,6} will be considered below as reliable.\\
The results of experiment PS--170 on the study of annihilation
$p \overline p \rightarrow e \overline e$ at low energies [10] have no
adequate interpretation till the present day. They resulted in an unexpected
behaviour of the proton electromagnetic form factor near the $p \overline p$--
threshold in the time--like region, where $s < 4.2~GeV^2$. The data on\\ $\mid
G\mid =\mid G_{m,p}\mid=\mid G_{e,p} \mid$ point to a large negative derivative
at the threshold that rapidly grows to zero or even to positive values at
$s\sim $  $4~GeV^{2} $. The magnitude of the derivative at the threshold is
determined by the threshold value $\mid G\mid =0.53\pm 0.05$. One of the early
values, $\mid G\mid =0.51\pm 0.08$, does not contradict the results of ref.
[10].  It was obtained [11] from the ratio of frequencies of $p \overline p$
annihilations at rest into $e\bar e$ and $\pi^+\pi^-$ pairs in liquid
hydrogen.  The determination of $\mid G\mid$ at the threshold is a complicated
problem since one should simultaneously consider the Coulomb and strong
interactions in the $p \overline p$--system and requires some approximations.
These approximations have been analysed in ref. [12] where a new scheme is
proposed for the determination of $\mid G \mid$. This scheme gave the value
$\mid G\mid =1.1$ that confirms the results of ref. [10].  Quite recently, a
new attempt has been undertaken for determining $\mid G\mid $ at the threshold
[13].  Combining the data on widths of $p \overline p$--atoms obtained in the
synchrotron trap with the results on the low--energy annihilation cross
section in $p \overline p$--system, the authors concluded that $\mid G\mid
=0.39$ or even $\mid G\mid =0.3$. This allows us to infer that there is no
abrupt change of $\mid G\mid$ at the threshold. Thus, the authors of [12,13]
propose a new view on the method of calculating $\mid G\mid$ at the threshold
from experimental data.\\
Let us now proceed to works that suggest the interpretation of the results
of the experiment [10]. In ref. [14] an attempt is made to consider the
interaction in the final state. The basic result is the formula
$G=ce^{i\delta}$, where $c$ is a slow variable function of $q^2$ at the
threshold ($q$ is the momentum in c.m.s. of the $p \overline p$--system) and
$\delta$ is the $N \overline N$ scattering phase. Since the phase $\delta$ is
complex at the threshold, we have \begin{equation} \mid G\mid =\mid c\mid
\cdot \mid 1-q\cdot\mbox{Im a}\mid~, \end{equation} where
$\mbox{a}$ is the complex scattering length. Owing to $\mid G\mid$ being
linear in $q$, the quantity $d\mid G\mid /ds$ is infinite at the threshold.
Analysis of the first four points from [10] with respect to the
$\chi^{2}$--criterion gives the values:  $\mid c\mid=0.53\pm 0.02\> ,\> \mbox{
Im a}=0.62\pm 0.08\>fm,\> \chi^{2}=0.07 $.  The authors of [14] employ the
values:$\mid c\mid =0.52,\mbox{Im a}\cong0.8\>fm$; they identify $\mbox{Im} a$
with the quantity $\mbox{Im a}(^{3}S_{1})$ computed from the experiment [15].
The description is qualitative since  $\chi^{2}\sim 10$.  The authors of [16]
assert that a good description of all known data on nucleon electromagnetic
form factors, including the data of [10], is obtained on the basis of a new
formulation of the vector--dominance model (VDM) and its subsequent
unitarization. In what follows, we will use different models of that type,
therefore we consider them in detail. They are based on the expressions for
the Dirac and Pauli nucleon form factors in VDM:  \begin{equation}
F_{N}(s)=\sum_{v} \frac{f_{v,NN}}{f_{v}}\frac{m^{2}_{v}}{m^{2}_{v}-s},
\end{equation} where $m_v$ is the mass of a vector meson, $f_{vNN}$ is the
coupling constant of a vector meson with a nucleon, $f_v$ is the universal
constant in the so--called identity of current and field.  Imposing
constraints on the parameters of formula (2), one can easily find the
experimental value $F_{N}(s=0)$ and the asymptotics following from the quark
counting rules [17] that coincides with the QCD--asymptotics within the
logarithmic accuracy. Then, the model is unitarized with the help of
a uniformizing variable. As a result, vector mesons acquire widths, and the
form factors can be calculated for all $s$.
So, all experimental data can be described both in the space--like ($s < 0$)
and time--like ($s > 0$) regions. Satisfactory description of more than
three hundred values of $\mid F_{N}\mid$ requires about ten free parameters in
the formula (2).  Besides, this approach
allows a model--dependent reproduction of the form of $\mbox{Im}F_{N} $,
$\mbox{Re}F_{N}$ in the whole time--like region. This fact will be used below.
Results of the analysis according to this scheme are presented in ref. [16].
The data of the experiment PS--170 are explained by including the third radial
excitation $\rho(770)$ with the mass $\sqrt{s}=2.15~GeV$ into formula (2) and
are plotted in Fig.1.
\section{Formulation of the analytical model}
It is easy to see that the nucleon form
factor, according to formula (2), has the following imaginary part
\begin{equation}
\mbox{ImF}_{N}=\pi\sum_{v} m_{v}^{2}\frac{f_{vNN}}{f_{\rho}}\delta(s-m_{v}^{2}).
\end{equation}
Formula (3) is an approximate expression obtained from the unitarity condition
which allows one to reproduce equation (2) with the use of dispersion relations
for $F_{N}$. We write the starting expression for the unitarity condition as
follows:
\begin{equation}
\mbox{Im}<o\mid j_{\mu} \mid N\bar N>=\sum_{n} <o\mid j_{\mu} \mid n><n\mid
T^{+}|N\bar N>,
\end{equation}
where $j_{\mu}$ is the electromagnetic current of a nucleon
$N$, and $|n>$ is the complete set of admissible intermediate states. In our
case, it is of the form
\begin{equation}
|n>=|2\pi >,\> |3\pi >,...,|K\bar K>,\> |N,\bar N>.
\end{equation}
Frazer and Fulco [18] were the first who
computed the contribution of the two--pion state and predicted the
$\rho$--meson on the basis of data on $F_{N}$. By choosing different terms in
the sequence (5), one can obtain many models of the type (2). Earlier, the
model of ref.[19] was used in [20] and the contribution of an $N \overline N$
intermediate state was calculated. This contribution is important for two reasons.
First its consideration results in a new branch point in formula (2), the
threshold of the reaction $N \overline N$ situated on the lower edge of the energy
region studied in ref.[10]. Second bound states or resonances in $N \overline N$ --system near the threshold will influence the behaviour
of $F_N(s)$ in the nonobservable region below the $N \overline N$--threshold
and in the observable region above the $N \overline N$--threshold investigated
in ref.[10]. It is clear that the state $|N \overline N>$ appears on the
background of the sum of other states of the series (5) and the result
is model--dependent.  Therefore, it is important to study the degree of
that dependence by considering another model differing from the one used in
[20] for $F_{N}(s)$ as a background for the state $|N \overline N>$. \\We will
take the model of ref.[21] formulated in terms of the Sachs form factors $G$
measured experimentally. The model is based on the formulae
\begin{eqnarray}
G_{m,p}(s)=\sum_{k=1}^{3}\frac{\epsilon_{k}(s)}{s-a_{k}-\gamma_{k}\sqrt{s_{k}-s}}\>,\quad
G_{e,p}(s)=\sum_{k=1}^{3}\frac{\beta_{k}(s)}{s-a_{k}-\gamma_{k}\sqrt{s_{k}-s}}\>, \\
\mbox{where}\quad
\epsilon_{k}(s)=\frac{\epsilon_{k}^{1}+\epsilon_{k}^{0}s}{s-a_{k}-\gamma_{k}\sqrt{s_{k}-s}}\>,\quad
\beta_{k}(s)=\frac{\beta_{k}^{1}+\beta_{k}^{0}s}{s-a_{k}-\gamma_{k}\sqrt{s_{k}-s}} \cdot
\end{eqnarray}
The energy behaviour of electromagnetic form factors is explained with the use
of three resonances:$\rho $, $\omega$, $\varphi $ specified by indices
$k=1,2,3$ in formula (6).  The masses, widths and thresholds
$a_{k},\gamma_{k},s_{k}$ are taken from experiment.  The model parameters are
the coupling constants
\begin{eqnarray}
(\beta_{1}^{1}+\epsilon_{1}^{0}s)f_{1}(s)=g_{\gamma \rho }(s)g_{\rho NN}(s),
\nonumber \\
(\beta_{2}^{1}+\epsilon_{2}^{0}s)f_{2}(s)=g_{\gamma \omega }(s)g_{\omega NN}(s),
\nonumber \\
(\beta_{3}^{1}+\epsilon_{3}^{0}s)f_{3}(s)=g_{\gamma
\phi }(s)g_{\phi NN}(s)\>, \nonumber \\
\mbox{ where} \nonumber\\
f_{k}(s)=\frac{1}{s-a_{k}-\gamma_{k}\sqrt{s_{k}-s}} \cdot
\end{eqnarray}
This unusual form of the constants is chosen by
analogy with the index of refraction in optics. They are not only
energy--dependent, but also contain a complex component when $s>s_{k}$.  The
coupling constants are chosen so as to be consistent with the known
experimental data at $s = 0$. Then, we are left with two free parameters $\epsilon_{2}^{0}$
and $\epsilon_{3}^{0}$ to be defined from the conditions required at $s \rightarrow
\infty$. The $SU(3)$ symmetry should hold on the asymptotics identically. This
condition seems to be the weakest one since it can be changed by including new
vector mesons into consideration. Therefore, the parameters $\epsilon_{2}^{0}$
and $\epsilon_{3}^{0}$ are determined according to the $\chi^{2}$ criterion on
the basis of experimental points $|G_{p}|$ cited in refs. [22].
An interesting feature of the model [21] is that it correctly describes
the ratio $|G_{p}|/|G_{n}|$ above the $p\bar p$--threshold. More exactly, it reproduces 
the experimental value $|G_{n}(s=4~GeV^{2})|=0.42\pm 0.06$ (see [23]). The model result for $|G_{p}|$ is drawn in Fig.2. and
$\epsilon^{0}_{2}=-3.41,\epsilon^{0}_{3}=3.23,\chi^{2}=10.1$. \\The influence of
the $|N\bar N>$~~ contribution to the unitarity condition (4) on
$|G|$ is computed in the same way as in refs. [20, 24]. We construct the analytic model for the forward elastic scattering amplitude $T$  in
terms of the uniformizing variable \begin{equation} z=\sqrt{\frac{4(s-\alpha
)}{s(4-\alpha )}}-\sqrt{\frac{\alpha (s-4)}{s(4-\alpha )}}, \end{equation}
where $s$ is the conventional Mandelstam variable equal to the square of the
total energy of a $p\bar p$--system in the c.m.s. in units $M_{p}$. The
variable $z$ contains branch points at $s = 0; 4$ corresponding to the
reaction threshold of elastic $pp$ and $p\bar p$--scattering and an effective
branch point at $s = \alpha$ corresponding to the nonobservable region for the
elastic $p\bar p$-- scattering.  The threshold of process $p\bar p\rightarrow
p\bar p$ is mapped into points $z = \pm1$ on the $z$--plane; whereas the infinit 
$s$--plane point, into points $\pm z_{1},\> \pm 1/z_{1}$, where 
$z_{1}=\sqrt{\frac{2-\sqrt{\alpha}}{2+\sqrt{\alpha}}}$.  Disposition of all the
four sheets of the Riemann surface of the function $z(s)$ is drawn in Fig.3
for $ \alpha=1.44$. In ref. [24] it is shown that the experimental data on
$\rho=\mbox{Re}T/\mbox{ImT}$ and $\sigma_{tot}$ can be well described
provided that the $p\bar p$--system possesses a quasinuclear bound state with
the binding energy $E=(1.88\pm 0.05)~MeV$ and width $\Gamma =(1.6\pm 0.1)~MeV$.
The scattering amplitude was taken in the form \begin{equation} T
=T_{b}+\frac{c_{\rho}}{z-(z_{\rho})_{1}}-\frac{c_{\rho}}{z-(z_{\rho})_{2}},
\end{equation}
where $T_{b}(s)$ is a
polynomial in $z$, $(z_{\rho})_{1,2}=1\mp \gamma\pm i\delta
$ and $\alpha =1.44, 10^2\gamma=-0.54\pm0.02, 10^2\delta=2.6\pm0.08$. The
pole terms represent the contribution of the quasinuclear state; whereas the
polynomial determines the contribution of a nonresonance background of S,P and
D--waves.  Special attention was paid to the threshold value of the $T$ amplitude which
is complex [24].  The amplitude (10) well describes the
experimental data up to $4.4 ~GeV^2$ in terms of the variable
$s$. It is valid in the vicinity of $z= 1$ and has two poles in distinction to
the usual quantum mechanical amplitude. Appearence of the two poles in the variable z inctead of the one pole in the variable $q$ in the scattering amplitude $T$ is a
consequence of choosing z as uniformazing variable. Another important feature of
the formula (10) is the form of the pole term contribution to $\mbox{Im T}$ and $\mbox{Re
T}$.  The bound state (pole) contribution to the $\mbox{ImT}(p_l=80 ~MeV/c)$ is about 10\% of the total value $\mbox{Im T}_{p\bar p}$. On the
other hand, the bound state contribution to the $\mbox{ReT}$ is
larger then the background one and ensure the correct value $\rho$ (see Fig.
4,5).  Near the $p\bar p$--threshold the pole contribution to the unitarity
condition (4) becomes dominant, and thus, we will restrict ourselves to the
 pole approximation.  Quantum numbers of this state are unknown.  A detailed
scheme of calculation corresponds to the scheme by Frazer and Fulco [18] for
the contribution of different partial waves to $\mbox{Im}F_{N}$. In our case,
it gives that these states are either $^3S_1$ or $^3D_1$. Then, the unitarity
condition (4) is reduced to the Riemann boundary--value problem [25] that can
be solved (see Appendix).  Inside the ring containing the unit circle (Fig.3) the
solution is of the form \begin{equation}
G_{pol}=\frac{c(z)}{\prod_{i=1}^{2}(z-(z_{\rho})_{i})(z+(z_{\rho}^{*})_{i})},
\end{equation}
where $c(z)$ is an entire function within which the solution is determined.
Setting
$c(z)=c_{1}(z)\cdot(z^{2}-z_{1}^{2})(1-z^{2}z_{1}^{2})/(1-z_{1}^{2})^2$, we
can ensure the asymptotic behaviour of $G_{pol}$ at infinity.  Taking
advantage of $c_1(z)$ being arbitrary, we assume the solution to be of the form
\begin{eqnarray}
\lefteqn{G_{pol}(z)\frac{(1-z_{1}^{2})^{2}}{(z^{2}-z_{1}^{2})(z^{2}z_{1}^{2}-1)}=} & &
\nonumber\\
&& A_{1}\biggl\{\Bigl(\frac{1}{z-(z_{\rho})_{1}}-\frac{1}{z-(z_{\rho})_{2}}\Bigr)-
\Bigl(\frac{1}{z+(z_{\rho}^{*})_{1}}-\frac{1}{z+(z_{\rho}^{*})_{2}}\Bigr)\biggr\}+
\nonumber\\
&& +A_{2}\biggl\{\Bigl(\frac{1}{z-(z_{\rho})_{1}}+\frac{1}{z-(z_{\rho})_{2}}\Bigr)-
\Bigl(\frac{1}{z+(z_{\rho}^{*})_{1}}+\frac{1}{z+(z+(z_{\rho}^{*})_{2}}\Bigr)\biggr\}.
\end{eqnarray}
Around the $p\bar p$--threshold the equalities $\mid G_{e,p} \mid=\mid
G_{m,p} \mid=\mid G \mid$ hold valid and, under this assumption, the experiment
in [10] was analysed.  Therefore, we put 
\begin{equation}
G_{e,p}+G_{m,p}=2G_{w},
\end{equation} 
where the
functions $G_{e(m),p}$ are given by formulae (6).  Considering the contribution
of the $|N\bar N>$--state to the unitarity condition (4), we obtain for the
proton electromagnetic form factor $G$:  
\begin{equation} 
G=G_{w}+G_{pol}.
\end{equation} 
We shall assume the position of poles to be known from ref.
[24]; then, the form factor $G$ depends on two free parameters $A_{1}, A_{2}$.
The behaviour of $G_{pol}$ on the upper edge of the cut $[\alpha, \infty)$
around the $N\bar N$--threshold is determined by the poles $(z_{\rho})_{1} $
and $(z_{\rho})_{2}$; whereas on the lower edge, by the poles
$(z_{\rho}^{*})_{1}$ and $(z_{\rho}^{*})_{2}$. If we calculate the common
denominator for the contributions of the poles $(z_{\rho})_{1}$ and $(z_{\rho})_{2}$
in the formula (12), the energy factor $(z - 1)$ will arise in front of the
parameter $A_2$; whereas a constant, in front of the parameter $A_1$.  This
allows us to draw analogy between the parameter $A_{1}$ and $\epsilon
^{1}_{k},\beta^{1}_{k}$ as well as between $A_{2}$ and $\epsilon
^{0}_{k},\beta ^{0}_{k}$ in formula (7).  The expression for $G_{pol}$ (11)
follows from the unitarity condition and analytic properties of the proton
form factor and $N\bar N$--scattering amplitude.  Therefore, formulae (6) are
substantiated, irrespective of the above  mentioned analogy with optics.
The result of the analysis (Fig. 4) according eq. (12) is presented in the
table 1 and parameters are equal:
$\alpha=0.23\pm0.04$, $\epsilon^o_2=2.97\pm0.03$, $\epsilon^o_3=3.23$,
$10^2A_1=0$, $10^2A_2=1.2\pm0.01$. 

\section{Discussion of the results}
 The parameters $A_1$ and $A_2$ representing the
coupling constants of a quasinuclear bound state are sensitive to the
background shape in formula (14) as follows  from comparision of this fit
 and the fit of ref.[20] ($A_{1}\not=0$ in ref.[20]). The magnitude of the
 background is determined by the parameters $\epsilon^{0}_{2}$ and
$\epsilon^{0}_{3}$ and is slow changing function in the s interval under
investigation.  The parameters $A_{1},A_{2},\alpha$ determine the rapid
change of $G$ in formula (14).  Via separating the parameters into these two
groups, we can obtain their statistically reasonable values (table 1).  The
analysis would be considerably simplified if the experimental values of
$s>4M^{2}_{p}$  were known for $\mbox{Im}G$ and $\mbox{Re}G$.  Their
determination requires polarization experiments whose theoretical study is
carried out in ref. [26]. \\ Recently two independent experiments gave new
 information on the $p\bar p$ interaction at low energy. The value of the
 $p\bar p$ annihilation total cross section down to the momenta  43~MeV/c
have been measured by OBELIX experiment [27] at LEAR and no resonant behaviour
of the cross section was found.The existence of some stracture in the $e\bar
e\to hadrons$ cross section near the $P\bar P$ threshold was indicated in
FENICE at ADONE [28]. A combined analysis of these data and the data on the
proton form factor provides a good candidate for the quasinucler bound state
with the mass $M=1.85\pm 0.01~GeV^{2}$ and the width $\Gamma=40\pm10~MeV$.
This candidate dosn't contradict our candidate [20].  Then the question arise
why this candidat is not seen in the OBELIX experiment on the $p\bar p$
annihilation cross section at very low energy.  The first reason for that is
the mass of 'baryonium' which is less then $2M_{p}$. The second is based on our
analytical model. In this model $(\sigma_{tot})_{pole}\simeq 0.1~\sigma_{tot}$
at low energy (see Fig. 4) but $\sigma_{ann}<\sigma_{tot}$.  From these
 inequalities it is clear why 'baryonium' is not seen in OBELIX data. On the
other hand, in FENICE experiment cross section $e^+e^-\rightarrow hadrons$
depends not only on $\mbox{ImT}$ but also on $\mbox{ReT}$ for 
which the pole contribution is large.  That is the reason why 'baryonium' is not
seen in OBELIX and seen in FENICE.  Thus the results of bouth this experiments
are consistent. \\ Finaly we mention a pure theoretical result; the
 method of derivation of formula (11) for describing a quasinuclear state can
be applied to any vector meson in formula (2). Therefore any vector meson
will be characterized not only by the mass and width but also by two
parameters like coupling constants.  In other words, the effective coupling
constants will be energy--dependent, what is assumed in ref. [21] and is
reflected in formulae (7).  

\section*{Appendix} The unitarity condition (4) is
an exact equation if use is made of the complete system of admissible
intermediate states (5), otherwise it is an approximate equation dependent on
the assumptions made. Let us take it in the form \begin{displaymath}
\mbox{ImF}=F(e^{i\delta}\sin {\delta})^{*}+\bar g, \end{displaymath} where
$\delta$ is the $N\bar N$--scattering phase with quantum numbers of the pole
state  unknown yet; $\overline g$ is the contribution of all other processes
in the same channel.  We reduce it to the form 
\renewcommand{\theequation}{A.\arabic{equation}}
\setcounter{equation}{0}
\begin{equation}
F=e^{2i\delta}F^{*}+2ig.
\end{equation}
The relation (A.1) is valid for $\mbox{Im}s=0$ and $\mbox{Re}s\geq 4M^{2}_{p}$.
The function $F$ is analytic in the complex plane $s$ with the cut $[4
M^2_{p}, \infty)$ outside of which $F^{*}(s)=F(s^{*})$.  This relation
represents a linear inhomogeneous Riemann boundary--value problem for the
function $F$. If $e^{2i\delta}$ has a pole near the cut, then in its
vicinity we can consider the homogeneous problem
 \begin{displaymath}
F=e^{2i\delta}F^{*}.
\end{displaymath} 
As it is known [25], the main difficulty in solving it consists in constructing a function  
analytic in the
plane $s$ and coincident on the cut with $e^{2i\delta}$.  However, if
$e^{2i\delta}$ is taken in the form admitting the analytic continuation onto
complex $s$, the problem is reduced to the solution of a functional equation
for $F$ in the uniformizing variable $z$. We will represent $e^{2i\delta}$ in
the form 
\begin{displaymath}
e^{2i\delta}=\prod_{j}\frac{(z-z_{j}^{*})(z+z_{j})}{(z-z_{j})(z+z_{j}^{*})}.
\end{displaymath}
The function $e^{2i\delta}$ is real on the
imaginary axis $z$, i.e. on the real axis $s$ when $s < \alpha$.  Equation
(A.1) is valid on the cut $[4M^2_{p}, \infty)$ that transforms into the real
axis $z=x+iy$ , and $F(s)\rightarrow F(x),\> F^{*}(s)\rightarrow F(-x)$
\begin{displaymath}
F(x)=\frac {(x-z_{j}^{*})(z+z_{j})}{(x-z_{j})(z+z_{j}^{*})}F(-x),
\end{displaymath}
where
we took only one pole, without loss of generality. The latter functional
equation for $F(x)$ can be written as follows
\begin{displaymath}
F(x)(x-z_{j})(x+z_{j}^{*})=G(x),
\end{displaymath}
\begin{displaymath}
G(x)=G(-x)
\end{displaymath}
and thus $F(z)$ is
representable in the form
\begin{displaymath}
F(z)=\frac{G(z)}{\prod_{j}(z-z_{j})(z+z_{j}^{*})},
\end{displaymath}
where $G(z)$ is an entire even function of the
variable $z$. The inhomogeneous boundary--value problem (A.1) can be solved
in a similar manner and formula (11) can be proved.
 \newpage 
\begin{table}
\centering \caption{} \vspace{0.7cm} \begin{tabular}{||c|c|c|c||} \hline $S\>
~GeV^{2}$ & $G_{exp}$ & $|G|$ & $\chi^{2}_{i} $ \\ \hline 3.523 & $ 0.53\pm
0.05$  &0.63 &3.9  \\ $3.553$& $0.39\pm 0.05$ &0.35 &0.63 \\ $3.57$ & $
0.34\pm 0.04$ &0.32 & 0.26 \\ $3.59$ &$0.31\pm 0.03$ &0.3 &0.15 \\ $3.76$ &
$0.26\pm 0.014$ &0.27&0.66 \\ $3.83$&$0.25\pm 0.01$ &0.27&1.9 \\ $3.94
$&$0.247\pm 0.014$&0.254&0.23 \\ $4.18$&$0.252\pm 0.011$&0.221&8.1 \\ \hline
\end{tabular} \end{table} \hspace{2.in} \newpage \begin{center} {\Large\bf
Figure Captions.} \end{center} Fig. 1. The curve from Fig. $3^a$ of ref.[16] on a
larger scale.  The quality of the fit PS-170 data is very poor.  \\ Fig. 2. Our
fit to the old data [22] by $G_w$.\\ Fig. 3. Disposition of four sheets of the
Riemann surface of the function z(s)\\for $\alpha=1.44$. The threshold $p\bar
p$ is mapped into points $z=\pm 1$.\\ Fig. 4. The pole contribution to the
$\mbox{ImT}.$\\Fig. 5. The pole contribution to the $\rho.$\\Fig. 6. Our fit to
PS-170 data with account of the pole contribution (eq.(12)). \newpage
\begin{figure}
\begin{center}
\includegraphics*[width=\textwidth]{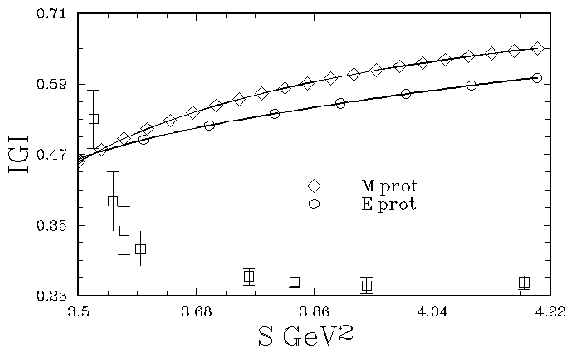}
\end{center} 
\caption{} 
\end{figure}
\newpage
\begin{figure}
\begin{center}
\includegraphics*[width=\textwidth]{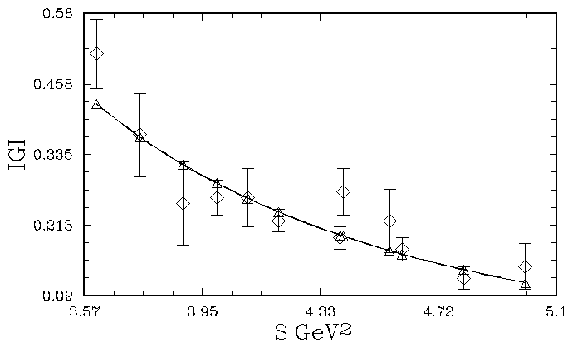}
\end{center}
\caption{}
\end{figure}
\newpage 
\begin{figure}
\begin{center}
\includegraphics*[width=\textwidth]{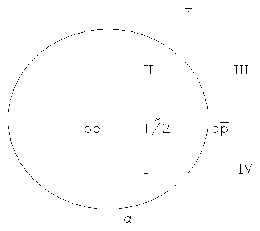}
\end{center} 
\caption{}
\end{figure}
\newpage
\begin{figure}
\begin{center}
\includegraphics*[width=\textwidth]{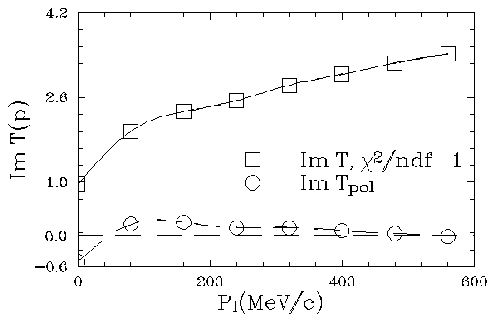}
\end{center}
\caption{}
\end{figure}
\newpage
\begin{figure}
\begin{center}
\includegraphics*[width=\textwidth]{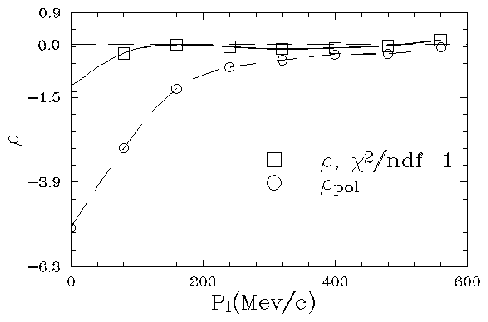}
\end{center}
\caption{}  
\end{figure}
\newpage
\begin{figure}
\begin{center}
\includegraphics*[width=\textwidth]{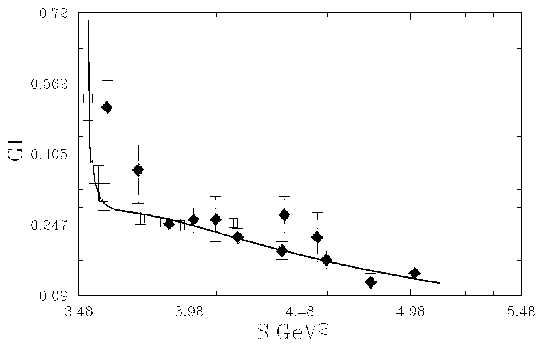}
\end{center}
\caption{}
\end{figure}
\end{document}